\documentclass[a4paper,11pt]{article}
\pdfoutput=1 % if your are submitting a pdflatex (i.e. if you have
\usepackage{jheppub} % for details on the use of the package, please
\usepackage[T1]{fontenc} % if needed

\usepackage{amsmath}
\usepackage{amssymb}
\usepackage{setspace}
\usepackage{bbm}
\usepackage{dsfont}
\usepackage{graphics}
\usepackage{hyperref}
\usepackage{comment}

\hypersetup{
    bookmarks=true,%
    colorlinks,%
    citecolor=blue,%
    filecolor=blue,%
    linkcolor=blue,%
    urlcolor=blue
}

\usepackage[font=footnotesize,labelfont=bf,justification=centerlast,width=.94\textwidth]{caption}

\usepackage{color}

\usepackage[normalem]{ulem}
\usepackage{epsfig}

\newenvironment{defn}[1][Definition]{\begin{trivlist}
\item[\hskip \labelsep {\bfseries #1}]}{\end{trivlist}}

\renewcommand{\bar}{\overline}
\renewcommand{\tilde}{\widetilde}

\renewcommand{\leq}{\leqslant}
\renewcommand{\geq}{\geqslant}

\newcommand{\Tr}{\operatorname{Tr}}
\newcommand{\nn}{\nonumber}

\newcommand{\const}{\operatorname{const}}

\newcommand{\UU}{\operatorname{U}}

\newcommand{\SL}{\operatorname{SL}}

\newcommand{\RR}{\mathbb{R}}

\newcommand{\ZZ}{\mathbb{Z}}

\newcommand{\calD}{\mathcal{D}}
\newcommand{\calE}{\mathcal{E}}

\newcommand{\calO}{\mathcal{O}}

\newcommand{\baa}{\bar a}
\newcommand{\eff}{\text{eff.}}
\newcommand{\vac}{\text{vac}}

\newcommand{\be}{\begin{equation}}
\newcommand{\ee}{\end{equation}}
\newcommand{\bea}{\begin{eqnarray}}
\newcommand{\eea}{\end{eqnarray}}

\newcommand{\p}{\partial}
\newcommand{\ba}{\begin{aligned}}
\newcommand{\ea}{\end{aligned}}

\usepackage[textsize=footnotesize,textwidth=1.5cm]{todonotes}

\newcommand{\YG}[2]{\textcolor{red}{#1}\todo[color=pink]{YG: #2}}

\title{\boldmath A note on the complex SYK model  and warped CFTs}

\author[a]{Pankaj Chaturvedi,}
\author[b]{Yingfei Gu,}
\author[a]{Wei Song,}
\author[a]{Boyang Yu}

\affiliation[a]{Yau Mathematical Sciences Center, Tsinghua University,\\ Beijing, 100084, China}
\affiliation[b]{Department of Physics, Harvard University,\\ Cambridge, MA 02138, USA}

\emailAdd{cpankaj@mail.tsinghua.edu.cn}
\emailAdd{yingfei\_gu@g.harvard.edu}
\emailAdd{weisong2014@mail.tsinghua.edu.cn}
\emailAdd{yuby16@mails.tsinghua.edu.cn}

\abstract{We discuss the connections between the complex SYK model at the conformal limit and  warped conformal field theories. Both theories have an $\operatorname{SL}(2,\mathbb{R}) \times \operatorname{U}(1)$  global symmetry. We present  comparisons on symmetries, correlation functions, the effective action and the entropy formula.
We also use modular covariance to reinterpret results in the complex SYK model.}

\begin{document}
\maketitle
\flushbottom

\section{Introduction and summary}

Symmetry plays an important role in investigating dualities between different theories. By analyzing the asymptotic symmetries, the seminal paper of Brown and Henneaux \cite{Brown:1986nw} predicted a connection between asymptotic Anti-de Sitter spacetime in three dimensions (AdS$_3$) and two dimensional conformal field theory, long before the discovery of the Gauge/Gravity correspondence\cite{Maldacena:1997re,Gubser:1998bc,Witten:1998qj}.  Similar ideas have also been applied to astrophysical black holes, the Kerr black holes, leading to the Kerr/CFT correspondence \cite{Guica:2008mu}. The Kerr/CFT correspondence has been subsequently generalized to various types of extremal black holes, see \cite{Bredberg:2011hp, Compere:2012jk} for a review. An essential input for Kerr/CFT is the near horizon geometry of extremal black holes or the so-called NHEK geometry, which is a fibration over a two dimensional Anti-de Sitter spacetime (AdS$_2$) \cite{Bardeen:1999px}. The isometry group for NHEK contains $\SL(2,\mathbb{R})\times \UU(1)$, where the $\SL(2,\mathbb{R})$  is the isometry of the AdS$_2$ factor.  Depending on the choice of boundary conditions, the asymptotic symmetry analysis for spacetimes with $\SL(2,\mathbb{R})\times \UU(1)$ isometry  indicates that the holographic dual is either organized by two dimensional conformal symmetry \cite{Guica:2008mu,Compere:2014bia} or the so-called warped conformal symmetry \cite{Detournay:2012pc}.  In either scenario, the Bekenstein-Hawking entropy of extremal and near extremal black holes can be interpreted as a Cardy \cite{Guica:2008mu} or Cardy-like formula \cite{Detournay:2012pc} and the greybody factors can be interpreted as thermal correlation functions of the putative dual field theory \cite{Bredberg:2009pv, Castro:2010fd,Chen:2010xu,Song:2017czq}. Furthermore,  logarithmic corrections and the $1$-loop partition function have also been computed in \cite{Pathak:2016vfc, Castro:2017mfj}. In this note, we would like to focus on the case where the $\SL(2,\mathbb{R})$ is enhanced to a Virasoro symmetry and the $ \UU(1)$ isometry to a $\UU(1)$ Kac-Moody symmetry \cite{Compere:2008cv,Compere:2013bya}. A two dimensional quantum field theory invariant under the Virasoro-Kac-Moody symmetry is called a warped conformal field theory (WCFT) \cite{Detournay:2012pc}.

Despite this progress, not many details are known about the putative field theory duals for Kerr black holes. To proceed, one might choose a top down approach, namely to embed Kerr black holes into string theory \cite{Detournay:2010rh, Guica:2010ej,Compere:2010uk, ElShowk:2011cm, Song:2011ii,Azeyanagi:2012zd} and study the irrelevant deformations \cite{ElShowk:2011cm} from more familiar examples of holography such as the D1-D5 black holes \cite{Strominger:1996sh}. Recent developments on integrable irrelevant deformations \cite{Guica:2017lia, Bzowski:2018pcy, Chakraborty:2018vja, Apolo:2018qpq} might help us improve our intuition. Alternatively, a bottom-up approach is to study simple models that capture some universal features. Two dimensional gravity theories on AdS$_2$ are natural candidates for this purpose. From the higher dimensional analysis, one expects the asymptotic symmetry to contain a conformal symmetry \cite{Strominger:1998yg} or a Virasoro-Kac-Moody symmetry \cite{Hartman:2008dq,Castro:2008ms,Castro:2014ima}. However, it is difficult to find non-vanishing central charges for AdS$_2$ gravity directly. A partial reason is that the near horizon geometry itself is non-dynamical \cite{Amsel:2009ev} and it does not allow finite energy excitations \cite{Maldacena:1998uz}. Recently the investigation of the Sachdev-Ye-Kitaev (SYK) models 
\cite{
Sachdev:1993PhRvL,
Parcollet:1999oa,
Kitaev:2015tk} and their connection to two dimensional gravity 
\cite{
Sachdev:2010um,
Almheiri:2014cka,
Sachdev:2015efa,
Maldacena:2016hyu,
Maldacena:2016upp,
Engelsoy:2016xyb,
Kitaev:2017awl} has opened a new paradigm within the framework of gauge/gravity duality. It turns out that in order to capture the universal physics of near extremal black holes,  it is important  to look at the region slightly further away from the horizon, instead of 
the near horizon throat itself. Correspondingly, the would-be conformal symmetry or warped conformal symmetry should be both explicitly as well as spontaneously broken.

The symmetry breaking pattern for conformal symmetry, namely its emergence in the IR and its subsequent breaking to $\SL(2,\mathbb{R})$ both explicitly and spontaneously, is captured by the  Jackiw-Teitelboim (JT) model \cite{Jackiw:1984je,Teitelboim:1983ux} and the SYK model.  The connection is sometimes called the nAdS$_2$/nCFT$_1$ correspondence to emphasize the deviation from pure AdS$_2$ in the bulk and the breaking of conformal symmetry  in the field theory side. The JT gravity is a two dimensional theory of dilaton gravity, which accommodates an exact solution with AdS$_2$ geometry. The conformal symmetry is explicitly broken by a running dilation. The SYK model is a $(0+1)$ dimensional quantum mechanical model involving $N\gg 1$ Majorana fermions with random all-to-all interactions. The model and its variants (for example see \cite{Gu:2016oyy,Witten:2016iux,Fu:2016vas,Klebanov:2016xxf,Gross:2016kjj,Davison:2016ngz, Krishnan:2016bvg,Turiaci:2017zwd, Murugan:2017eto,Chen:2017dav,Jian:2017unn,Cai:2017vyk,Peng:2017kro,Maldacena:2018lmt}) possess remarkable properties in the low temperature/infrared (IR) limit, such as solvability, approximate conformal invariance  and maximal chaos \cite{Maldacena:2015waa}.  The low energy effective actions of the SYK model and the JT model are both described by a Schwarzian derivative \cite{Maldacena:2016upp,
Engelsoy:2016xyb,
Almheiri:2016fws, 
Maldacena:2016hyu,
Jensen:2016pah} which governs the dynamics of a pseudo-Goldstone mode associated to reparametrization invariance. Moreover, the entropy difference between near extremal black holes and extremal ones is linear in the Hawking temperature, resembling the linear growth of entropy at low temperatures in the SYK model.

Going back to extremal black holes in higher dimensions, as we pointed out earlier, the near horizon isometry is $\SL(2,\mathbb{R})\times \UU(1)$ instead of $\SL(2,\mathbb{R})$. Moreover, the $\UU(1)$ isometry has played a crucial role in reproducing the zero temperature entropy \cite{Guica:2008mu,Detournay:2012pc}. While the previous higher dimensional analysis focused on the near horizon limit and thus the unbroken symmetries, we expect to see more dynamics with the symmetry slightly broken. For this purpose, we will consider the $\operatorname{SL}(2,\mathbb{R}) \times \operatorname{U}(1)$ symmetry, its enhancement to Virasoro-Kac-Moody symmetry and the subsequent broken phase. A quantum mechanical model which realizes a similar symmetry breaking pattern is the complex SYK model \cite{Sachdev:2015efa,Davison:2016ngz} which preserves $\UU(1)$ charge and is relevant for describing condensed matter systems with disordered metallic states without quasi-particle excitations. On the other hand, the unbroken symmetry is the defining property of the warped conformal field theories WCFTs \cite{Hofman:2011zj, Detournay:2012pc},  which are conjectured to be dual to gravity at the near horizon region of near extremal black holes under certain boundary conditions \cite{Detournay:2012pc, Compere:2013bya}.

In this work, we will connect the SYK model with complex fermions  to WCFTs.  The SYK model with Majorana fermions is sometimes viewed as a near CFT (nCFT). In the same spirit, we would like to propose that the complex SYK model is a realization of a near WCFT (nWCFT).  Our goals are to use WCFT to better organize calculations in the complex SYK model and, on the other hand, to use the complex SYK model to understand more details of quantum gravity for near extremal black holes. We will leave the bulk analysis to a future publication \cite{Chaturvedi:2018x}. See also \cite{Cvetic:2016eiv, Gaikwad:2018dfc,Anninos:2017cnw, Larsen:2018iou, Castro:2018ffi} for related discussions.

The main observations of this note are the following:
\begin{enumerate}
\item 
The low energy limit of the complex SYK model has an emergent reparametrization and time dependent phase shift symmetry, which is similar to the Virasoro-Kac-Moody symmetry in WCFTs. Moreover, there is a spontaneous and explicit symmetry breaking to $\SL(2,\RR)\times \UU(1)$ in the complex SYK  which can be compared to the same global $\SL(2,\RR)\times \UU(1)$ symmetry in WCFTs. The similar symmetry structure of the two sides also supports the same form of correlation functions.

\item 
The effective action of the complex SYK model arises from the explicit symmetry breaking induced by the UV perturbation. While in the WCFT side, the effect of the symmetry breaking is characterized by the conformal anomaly, quantified by central charge $c$ and level $k$. We observe that the effective action of the complex SYK model can be compared with the Virasoro generator in WCFTs.

\item 
The effective action of the complex SYK is defined on the imaginary time circle and the Virasoro generator is defined on the spatial quantization circle. This makes the comparison between  them a bit unnatural. However, when we consider the partition function of WCFTs, we will see that these circles are naturally related by modular transformations. Furthermore, we can evaluate the partition function of both sides including the fluctuations. In particular the vacuum character gives a $\beta^{-2}$ contribution to the partition function which matches the 1-loop correction in the complex SYK. 
\end{enumerate}
The layout of this note is as  follows, in section 2, we briefly review the complex SYK model and WCFTs. In section 3 we discuss the symmetries, correlation functions, the effective action and partition function of the complex SYK model from the perspective of WCFT. In particular we use modular covariance of the partition function to interpret the connections between the complex SYK and WCFTs.

\section{Basics}

In this section, we give a basic introductions of the complex SYK model and WCFTs. We elucidate more on the detailed properties and the connections between these two different theories in the next section.

\begin{defn}[The complex SYK model.]
The complex SYK model is a quantum mechanical model of $N\gg 1$ complex fermions with random all-to-all interactions.
 The model has a global $\UU(1)$ symmetry and can be described by the Euclidean action:
\begin{align}
S = \int d\tau\left[ \sum^{N}_{i=1} \psi^{\dagger}_{i}(\partial_{\tau}-\mu)\psi_{i} + \sum_{ \{ i_a \} } J_{i_1\ldots	 i_q} \psi^{\dagger}_{i_1} \ldots \psi^{\dagger}_{i_{\frac{q}{2}}}  \psi_{i_{\frac{q}{2}+1}} \ldots \psi_{i_q}\right].\label{actioncsyk}
\end{align}
Here, $q\geq 4$ is an even integer and $\mu$ is the chemical potential that can be tuned to change the  $\UU(1)$ charge $Q=\sum^{N}_{i}\left\langle\psi^{\dagger}_i \psi_{i}\right\rangle$.
The couplings $J_{i_1\cdots i_q}$ are Gaussian random complex numbers with zero mean and obey
\begin{align}
J_{i_1 \ldots i_{\frac{q}{2}} i_{\frac{q}{2}+1} \ldots i_q }= J^*_{ i_{\frac{q}{2}+1} \ldots i_q i_1 \ldots i_{\frac{q}{2}}} , \quad
\overline{|J_{i_1\ldots i_q}|^2} = \frac{ \left( \frac{q}{2}  \right)!^2}{q N^{q-1}}J^2,\label{cmj}
\end{align}
where the label set  $\{ i_a \} : = \{ 1 \leq i_1 < \ldots < i_{q/2} \leq N,~ 1 \leq i_{q/2+1} < \ldots < i_{q} \leq N\} $. The properties of the complex SYK model, such as the partition function, correlation functions and maximal chaotic behavior have been studied in \cite{Davison:2016ngz,Bulycheva:2017uqj,Bhattacharya:2017vaz}.

In the large $N$ limit, the model self-averages and the averaged effective action can be written in terms of bilocal fields $G(\tau_1,\tau_2)$, $\Sigma(\tau_1,\tau_2)$ as
\begin{align}
\frac{S}{N} = &- \Tr \ln (-\partial_{\tau}+\mu -\Sigma) \nn \\
& -\int^{\beta}_{0} d^2\tau \left[\Sigma(\tau,\tau')G(\tau',\tau)+\frac{(-1)^{\frac{q}{2}}J^2}{q} G(\tau,\tau')^{\frac{q}{2}} G(\tau',\tau)^{\frac{q}{2}}  \right].
\label{actioncsyk3}
\end{align}
Physically, $G(\tau,\tau')=- \frac{1}{N} \sum_{i=1}^N \langle  {\psi_i(\tau) \psi_i^{\dagger}(\tau')} \rangle $ represents the Green's function and $\Sigma(\tau,\tau')$ represents the self energy. The large $N$ saddle point equation for the fields $G$ and $\Sigma$ are
\begin{align}
G(i\omega_n)=\frac{1}{i\omega_n+\mu-\Sigma(i\omega_n)}, \quad
\Sigma(\tau)=-(-1)^{ \frac{q}{2}} J^2 G(\tau)^{ \frac{q}{2}} G(-\tau)^{\frac{q}{2}-1},\label{Gfunc}
\end{align}
where $\omega_n=\frac{2\pi}{\beta}(n+\frac{1}{2})$ denotes the Matsubara frequencies. This set of saddle point equations can be solved numerically. Specifically, in the low temperature limit $N \gg \beta J \gg1$, the solutions approach a conformal form which will be discussed in details in the next section.
\end{defn}

\begin{defn}[Warped conformal field theories.] \label{wcft}
Warped conformal field theories (WCFTs) are characterized by the warped conformal symmetry.
The global symmetry is
$\SL(2,\mathbb{R}) \times \UU (1)$, while the local symmetry is described by a
 Virasoro-Kac-Moody algebra \cite{Hofman:2011zj, Detournay:2012pc}.  WCFTs originates from exploring the holographic duals for warped AdS$_3$ spacetimes \cite{Anninos:2008fx, Compere:2008cv}, which can also be obtained from near extremal black holes with an $\operatorname{SL}(2,\mathbb{R}) \times \operatorname{U}(1)$ invariant near horizon geometry in higher dimensions \cite{Bredberg:2009pv, Guica:2010ej, Song:2011ii}.  Later it was realized that AdS$_3$ with Dirichlet-Neumann boundary conditions \cite{Compere:2013bya} also has warped conformal symmetries. 
Using modular properties of WCFT, a Cardy-like formula can be derived that provides a microscopic interpretation of the Bekenstein-Hawking entropy for black holes \cite{ Detournay:2012pc, Azeyanagi:2018har}.  Explicit models for WCFT have been constructed in \cite{Compere:2013aya, Castro:2015uaa,Jensen:2017tnb}.  Recently, holographic entanglement entropy \cite{Castro:2015csg, Song:2016gtd}, connections to Rindler holography \cite{Afshar:2015wjm}, correlation functions \cite{Song:2017czq}, and modular bootstrap \cite{Apolo:2018eky} have also been discussed in this context. 
 
 WCFTs are invariant under the following finite coordinate transformations,
\begin{align}
z'=f(z),\qquad \bar z'=\bar z+g(z),\label{wcftTrans}
\end{align}
where $z,\bar z$ are two independent coordinates, and $f(z)$ and $g(z)$ are two arbitrary functions of the $z$-coordinate. 
The corresponding transformations for the current operator and energy-momentum tensor $P(z)$, $T(z)$ are given by:
\begin{align}
{P}(z) &= f'(z) P'(f(z)) - \frac{k}{2} g'(z) \,, \nn \\
{T}(z)&= \left(f'(z)\right)^2 T'(f(z))+ g'(z) f'(z) P'(f(z)) - \frac{k}{4} \left( g'(z) \right) ^2  + \frac{c}{12} \{f(z), z \} \,,
\label{eqn: finiteTrans}
\end{align}
where $\{f(z),z\}$ is the Schwarzian derivative defined by
\begin{align}
\{f(z),z \} =\frac{f'''(z)}{f'(z)}  - \frac{3}{2} \left(
\frac{f''(z)}{f'(z)}
 \right)^2.
\end{align}
When defined on the plane the warped conformal symmetry is generated by a set of vector fields
\begin{align}
l_n=-z^{n+1}\partial_z,\quad p_n=-i z^{n}\partial_{\bar z},\label{wcftgen}
\end{align}
where the infinitely many conserved charges corresponding to these generators  can be written as
\begin{align}
L_n=-\frac{i}{2\pi}\int\mathrm{d}z~z^{n+1}T(z)~,
\quad P_n=\frac{1}{2\pi}\int\mathrm{d}z~z^nP(z)~.\label{wcftcharge}
\end{align}
These conserved charges form a warped conformal algebra \cite{Detournay:2012pc} which is nothing but a copy of the Virasoro algebra and a $\operatorname{U}(1)$ Kac-Moody algebra described by
\begin{align}
[L_n, L_m]=&(n-m)L_{n+m}+\frac{c}{12}n(n^2-1)\delta_{n, -m}\,,
\nonumber\\
[L_n, P_m]=&-mP_{n+m}\,,\nonumber\\
[P_n, P_m]=&k\frac{n}{2}\delta_{n, -m},\label{wcftVU1}
\end{align}
where $c$ is the central charge and $k$ is the Kac-Moody level \cite{Detournay:2012pc}. The above algebra has an $\SL(2,\RR)\times \UU(1)$ sub-algebra generated by $L_{\pm1}$, $L_0$ and $P_0$. Specifically, the conserved charges are defined in (\ref{wcftcharge}) such that $L_0$  and $L_{-1}$ generate scaling and translation in the $z$-direction whereas, $P_0$ generates translation in the $\bar z$-direction. The charges act on primary operators $\calO$ which are labeled by the conformal weight $\Delta$ and charge Q as,
\begin{align}
\left[{L}_n,\mathcal{O}(z,\bar z)\right]&=\left[ z^{n+1}\partial_z+(n+1)z^n\Delta\right]{O}(z,\bar z),\label{optrans1}\\
\left[{P}_n,\mathcal{O}(z,\bar z)\right]&=         {-}i z^n\partial_{\bar z} \mathcal{O}(z,\bar z)=z^n Q \mathcal{O}(z,\bar z).
\label{optrans2}
\end{align}
For a unitary WCFT, the charges $L_n$ and $P_n$ in Eq.(\ref{wcftcharge}) are hermitian and the spectrum satisfies
\bea
k>0,\quad Q\in \mathbb{R}, \quad  c\geq 1, \quad  \text{and}~\Delta\geq \frac{Q^2}{k}.\label{unitc2}
\eea
However, it has been noticed \cite{Compere:2008cv,Detournay:2012pc, Compere:2013bya, Apolo:2018eky} that unitary WCFTs are not holographically dual to gravitational theories with a metric formulation. Instead, holographic WCFTs feature a negative level, large central charge, and pure imaginary charge for the vacuum which is to say
\be
k<0,\quad c\gg1,\quad  i P_0^{\vac} \in \mathbb{R} \,.
\ee
Recently, the modular bootstrap has been applied to the WCFT in \cite{Apolo:2018eky}, where more constraints have been discovered. Interestingly, though holographic WCFTs are non-unitary, the vacuum module is still unitary.

Here we briefly comment\footnote{We thank Luis~Apolo for discussions. } on the connections among (1) a chiral half of a generic two dimensional CFT with an internal $\UU(1)$ symmetry, and (2) a chiral CFT which is a chiral sector of a holomorphically factorized 2d CFT with an internal $\UU(1)$ symmetry, (3) a WCFT, and (4) a chiral CFT from a DLCQ of a 2d CFT.  Although the algebra can appear to be similar in these theories, there are subtle differences:
\begin{enumerate}
\item A chiral sector of a generic two dimensional CFT can be embedded into Lorentz invariant theories.
In general, a 2d CFT does not factorize and the chiral sector is not modular invariant by itself.
The holographic dual also requires combining both the chiral and anti-chiral sectors.  Examples of the bulk dual can be Einstein gravity on AdS$_3$ spacetime with an extra gauge field \cite{Fitzpatrick:2015zha}.
In this regard, the focus in the literature has been on unitary theories with a positive Kac-Moody level, a real spectrum of charges, and a vacuum with zero charge.

\item A chiral CFT has central charge  $c_R=0$, $c_L=24n$, $n\in{\mathbb Z}$. Such theories are Lorentz invariant.  Modular invariance and holomorphicity impose strong constrains, and an operator is either a current or a descendant of a current. There are some rare examples of chiral CFTs with $c=24$ \cite{Schellekens:1992db} including the so-called monster CFT. Using the $c=24$ CFTs as seeds, one can build chiral CFTs with larger central charges.  As was discussed in \cite{Perlmutter:2016pkf}, a chiral CFT is not chaotic. It was suggested \cite{Witten:2007kt} that pure Einstein gravity with negative cosmological constant is holographically dual to the so-called extremal CFTs.  The characteristics of extremal CFTs are that they should be holomorphically factorizable and have the most sparse spectrum. Further discussions can be found in \cite{Maloney:2007ud,Benjamin:2016aww}. Purely chiral CFTs  are conjectured to be holographically dual to chiral gravities \cite{Li:2008dq, Maloney:2009ck}.

\item A WCFT is not Lorentz invariant. Virasoro-Kac-Moody primary Operators are not local in the $\UU(1)$ direction.
As was discussed in \cite{Song:2017czq, Apolo:2018eky}, several computational techniques can be borrowed from theory (i), namely 2d CFT with chiral $\UU(1)$s. However, modular properties are simpler in WCFTs.
While unitary WCFTs also exist, holographic WCFTs are non-unitary in the sense that the Kac-Moody level is negative. As was discussed in \cite{Apolo:2018eky}, holographic WCFTs also feature a spectrum containing pure imaginary charges. A non-vanishing vacuum charge is also essential in reproducing  black hole entropy \cite{Detournay:2012pc} and holographic entanglement entropy \cite{Castro:2015csg, Song:2016gtd}.

\item  Finally, as a digression, another version of chiral CFT can be obtained by a DLCQ procedure from a 2d CFT \cite{Balasubramanian:2009bg}. In the DLCQ limit, only the right-movers are allowed to be excited, and hence the theory is chiral and not Lorentz invariant. Moreover, \cite{Balasubramanian:2009bg} shows that such theories are also related to the Kerr/CFT correspondence \cite{Guica:2008mu}, where the $\UU(1)$ isometry of the near horizon geometry is enhanced to a Virasoro. This is in contrast to the case of WCFT which enhances the $\SL(2,\mathbb{R})$ isometry to Virasoro.
\end{enumerate}
\end{defn}

\section{Observations}

In this section we discuss the connections between WCFT and the complex SYK model at conformal limit (i.e. $N\gg \beta J \gg 1$), and  elucidate how the relevant properties of both theories are related to each other. We observe that the correlators, the effective action, and the one-loop corrections of  the complex SYK can be related to correlators, Viraroso generators and the vacuum character of WCFTs. 
 We propose that the complex SYK model can be considered as an example of near-WCFT.

In general, the complex SYK model can be related to any of the four candidate theories mentioned in the previous section with the symmetry breaking pattern: Virasoro-Kac-Moody to $\SL(2,\RR) \times \UU(1)$.
However, here our main focus is on the possible connection between the complex SYK and WCFTs.  Such connections are also important for understanding the relation of the complex SYK to the extremal or near-extremal black holes in the context of holographic duality.

\subsection{Symmetries and correlation functions}
Here we describe in detail the symmetries and the correlators of the two theories and show how they are related to each other.
\begin{defn}[The complex SYK model at the conformal limit.]
As discussed in \cite{Sachdev:2015efa},
in the low energy/IR regime ($\omega, T\ll J$) the Green's function and the self-energy (absorbing $\mu$ in frequency space) satisfy the following equations
\begin{align}
\int d\tau' \Sigma_c(\tau,\tau')G_c(\tau',\tau'')&=-\delta(\tau-\tau''), \nn \\
\Sigma_c(\tau,\tau')&=-(-1)^{\frac{q}{2}} J^2 G_c(\tau,\tau')^{\frac{q}{2}} G_c(\tau',\tau)^{\frac{q}{2}-1},\label{Gfunc3}
\end{align}
here the subscript $c$ stands for  ``conformal'' (IR solutions). The equations in (\ref{Gfunc3}) have the following symmetry
 \begin{align}
G_c(\tau_1,\tau_2)&\to G'_c(\tau_1,\tau_2)= [f'(\tau_1)f'(\tau_2)]^\Delta e^{i  (\Lambda(\tau_1)-\Lambda(\tau_2))}G_c(f(\tau_1),f(\tau_2)),\label{GSfuncC}\\ \Sigma_c(\tau_1,\tau_2)&\to \Sigma'_c(\tau_1,\tau_2)=[f'(\tau_1)f'(\tau_2)]^{1-\Delta}e^{i (\Lambda(\tau_1)-\Lambda(\tau_2))}\Sigma_c(f(\tau_1),f(\tau_2)),\label{GSfuncCS}
 \end{align}
where $\Delta=1/q$ is the conformal weight. The functions $f(\tau)$ and $\Lambda(\tau)$ represent the time reparametrizations and $\UU(1)$ fluctuations respectively.  At finite temperature $T=\beta^{-1}$ , the saddle point equations in (\ref{Gfunc3}) admit a solution:
\begin{align}
G_c(\tau_1,\tau_2)= \left\{
\begin{array}{ccc}
\displaystyle C_s e^{-\frac{2\pi{\cal E}}{\beta}(\tau_1-\tau_2)}\left(\frac{\pi}{\beta\sin  \frac{\pi}{\beta} |\tau_1-\tau_2|}\right)^{2\Delta},~~0<\tau_1-\tau_2 <\beta \\[1em]
\displaystyle -C_s e^{-\frac{2\pi{\cal E}}{\beta}(\tau_1-\tau_2)-2\pi{\cal E}}\left(\frac{\pi}{\beta\sin\frac{\pi}{\beta} |\tau_1-\tau_2|}\right)^{2\Delta},~~-\beta<\tau_1-\tau_2 <0
\end{array} \right.  \label{Gfinite}
\end{align}
where the coefficient $C_s$ is a normalization constant given by
\be
C_s= -\left( \frac{(1-2\Delta) \sin (\pi \Delta + \theta )  \sin (\pi \Delta - \theta ) }{ \pi  J^2 \sin (2\pi \Delta) }  \right)^{\Delta} \left(  \frac{\sin \left( \pi \Delta + \theta \right)}{\sin \left( \pi \Delta - \theta \right)} \right)^{\frac{1}{2}}
\ee
and the asymmetric factor $\theta$ is related to $\mathcal {E}$ by $e^{2\pi \mathcal E } = \frac{\sin (\pi \Delta + \theta)}{\sin (\pi \Delta - \theta)} $
and determined by UV details in the model.
  In the limit $\beta\to \infty$,  the expression for $G_c(\tau_1,\tau_2)$ in (\ref{Gfinite}) reduces to the zero temperature correlation function as
\bea
G_c (\tau_1,\tau_2) = \left\{
\begin{array}{ccc}
\displaystyle  \frac{C_s}{\pi |\tau_1-\tau_2|^{2 \Delta}} &,& \mbox{ $\tau_1-\tau_2 > 0$ } \\[1em]
\displaystyle - \frac{C_s e^{-2\pi \mathcal E}}{\pi |\tau_1-\tau_2|^{2 \Delta}} &,& \mbox{$\tau_1-\tau_2 < 0$}
\end{array} \right . \label{Gzero}
\eea
We emphasize here that the $\SL(2,\RR)\times \UU(1)$ symmetry acts on the solution (\ref{Gfinite}) in a slightly twisted way due to the non-zero $\mathcal{E}$\cite{Davison:2016ngz}.
 At $\calE=0$ (i.e. zero chemical potential)
  the $\SL(2,\RR)$ symmetry acts on the imaginary time circle through a M\"obius transformation given by
\begin{align}
\tan \frac{\pi}{\beta} f(\tau) = \frac{a \tan \frac{\pi}{\beta} \tau+b}{c \tan \frac{\pi}{\beta} \tau+d} ,\quad \begin{pmatrix}
a & b \\
c & d
\end{pmatrix}    \in \SL(2,\mathbb{R}).\label{pmatrix}
\end{align}
However, for non-zero $\calE$ the transformation $\tau \rightarrow f(\tau)$ will also affect the exponential factor and need to be absorbed by the extra $\UU(1)$ symmetry. Thus for $\mathcal E\neq 0$, the saddle point solution (\ref{Gfinite}) is invariant under a combined $\SL(2,\mathbb{R})$ and $\UU(1)$ transformation given by 
\begin{align}
\tan \frac{\pi}{\beta} f(\tau) &= \frac{a \tan \frac{\pi}{\beta} \tau+b}{c \tan \frac{\pi}{\beta} \tau+d} ,
\quad 
i\Lambda (\tau ) = \frac{2\pi \calE}{\beta} \left( f(\tau)-\tau \right)\,,
\label{sl2tr}
\end{align}
and an independent global $\UU(1)$ transformation
\begin{align}
\Lambda(\tau)\rightarrow \Lambda(\tau)+ \Lambda_0 
\,.
\label{u1tr}
\end{align}
\end{defn}

\begin{defn}[WCFTs.]
The structure of the two point correlator for WCFT is fixed by the warped conformal symmetry \cite{Song:2017czq,Castro:2015uaa}. Considering two primary operators ${\cal O}_1(z_1,\bar z_1)$ and ${\cal O}_2(z_2,\bar z_2)$
with conformal weights and charges as $(\Delta_1,Q_1)$ and $(\Delta_2,Q_2)$ respectively, the two point function on the vacuum state $|0\rangle$ is defined as,
\begin{align}
G_{W}(z_1,z_2;\bar z_1,\bar z_2)=\left\langle 0|T{\cal O}_1(z_1,\bar z_1){\cal O}_2(z_2,\bar z_2)|0\right\rangle,\label{2pf}
\end{align}
here $T$ stands for ordering in $z$ and we use the subscript $W$ to denote WCFT correlator. The two point function satisfies the following transformation rule
\be
G'_{W}(z_1,z_2;\bar z_1,\bar z_2)=\left(\frac{\partial z'_1}{\partial z_1}\right)^{\Delta_1}\left(\frac{\partial z'_2}{\partial z_2}\right)^{\Delta_2}G_{W}(z'_1,z'_2;\bar z'_1,\bar z'_2)
\,.
\label{2pf1}
\ee
On the neutral vacuum on the plane, two point functions are invariant under the global $\SL(2,\mathbb{R})\times \UU(1)$ symmetry and given by
\bea
G_{W}(z_1,z_2;\bar z_1,\bar z_2)=C\delta_{\Delta_1,\Delta_2}\delta_{Q_1,-Q_2}\frac{e^{i (\bar z_1-\bar z_2)}}{(z_1-z_2)^{2\Delta}},\label{2pf2}
\eea
where $C$ is some arbitrary constant. The above correlator is non zero only if the scaling dimensions of the two operators are the same (i.e. $\Delta_1=\Delta_2=\Delta$) and the 
 charges are opposite, for convenience we set $Q_1=-Q_2=Q_0=1$.

The two point correlator can also be obtained at finite temperature $T=\beta^{-1}$ and chemical potential $\bar{\mu}$
for the WCFT by going to the tilted cylinder \cite{Song:2017czq}, 
\begin{align}
z=e^{i \frac{2\pi }{\beta} x},
\quad
\bar z=y-
i\bar{\mu}x,
\label{finiteTr}
\end{align}
which introduces the thermal identification
\begin{equation}
(x,y) \sim (x+\beta,y +
i\bar\mu \beta) \,,
\label{thermalid}
\end{equation}
where $x$ stands for the Virasoro direction.
More explicitly, we obtain the fermion two point correlator by applying the transformation rule (\ref{2pf1}) on the zero temperature result as
\begin{align}
G_W (x_1,x_2;y_1,y_2) =
\left\{
\begin{array}{ccc}
\displaystyle C e^{i (y_1-y_2)+ \bar\mu (x_1-x_2)}\left(\frac{\pi}{\beta\sin \frac{\pi}{\beta} |x_1-x_2|}\right)^{2\Delta_{W}},~0<x_1-x_2<\beta \\[1em]
\displaystyle  -Ce^{i (y_1-y_2) + \bar\mu (x_1-x_2+\beta)}\left(\frac{\pi}{\beta\sin \frac{\pi}{\beta} |x_1-x_2|}\right)^{2\Delta_{W}},~-\beta<x_1-x_2<0
\end{array} \right . \label{2pffiniteE}
\end{align}
Furthermore, if we take the limit $\beta\rightarrow \infty$ with $\bar\mu\beta\equiv -2\pi\mathcal{E}$ fixed, we get the zero temperature correlator:
\bea
G_W (x_1,x_2;y_1,y_2) = \left\{
\begin{array}{ccc}
\displaystyle
\frac{C e^{i  (y_1-y_2) } }{\pi |x_1-x_2|^{2\Delta_W}}
,~~ x_1-x_2>0 \\[1em]
\displaystyle
- \frac{C e^{i (y_1-y_2)- 2\pi \calE } }{\pi |x_1-x_2|^{2\Delta_W}}
,~~ x_1-x_2<0
\end{array} \right . \label{2pfZero}
\eea
Alternatively, the above result (\ref{2pfZero}) can be obtained from the plane correlator (\ref{2pf2}) by a spectral flow transformation $ \bar z=y - i \bar\mu x$, 
as described in \cite{Song:2017czq}.
\end{defn}

\begin{defn}[The matching.]
The  discussion above exhibits many common features between the complex SYK model and WCFTs.
In particular, the emergent reparametrization symmetry in the conformal limit of the complex SYK can be interpreted as the Virasoro symmetry, while the phase fluctuation $\Lambda(\tau)$ can be associated with the  $\UU(1)$ Kac-Moody symmetry.  Thus, based on symmetry arguments, here we discuss two ways for identifying the coordinates such that one can relate the WCFT correlator to the complex SYK correlator.

\begin{defn}[Option I.] In this first approach we directly identify the Euclidean time $\tau$ (\ref{pmatrix}) as the Virasoro direction $x$ (\ref{thermalid}) and propose the map between correlation functions of WCFT and complex SYK model as follows
\bea
G_{W}(\tau_1,\tau_2;y_1,y_2)\leftrightarrow e^{i(y_1-y_2)} G_{c}(\tau_1,\tau_2) ,\label{GwGcmatch}
\eea
 with conformal weight $\Delta_W=\Delta$ and same $\UU(1)$ charge.
The mapping is natural from the perspective of symmetries.
Note that this identification depends on a particular   $\UU(1)$ direction, namely the $y$ direction, which is associated to a chemical potential $\bar\mu$. 
Both sides are invariant under the $\SL(2,\RR) \times \UU(1)$ transformaiton. More explicitly, the correlators in (\ref{GwGcmatch}) satisfy:
\begin{align}
(l_n^{(1)}+l_n^{(2)})G(x_1,x_2;y_1,y_2)&=0 \quad \text{for} \quad n=0 ~ \text{and} ~\pm1 \,, \\
(p_0^{(1)}+p_0^{(2)})G(x_1,x_2;y_1,y_2)&=0,
\end{align}
where the superscript indicates  the coordinates the operator acts on and the operators ${l}_n^{(a)}$
and $p_n^{(a)}$ are written on the tilted cylinder \cite{Song:2017czq} as
\be
l_n^{(a)}=i\frac{\beta}{2\pi}e^{i\frac{2\pi n }{\beta} x_a}(\partial_{x_a}
+
i\bar\mu \partial_{y_a})- n e^{i\frac{2\pi n }{\beta} x_a}\Delta,~~p_n^{(a)}={-}ie^{i\frac{2\pi n }{\beta} x_a}\partial_{y_a} \quad \text{for} \quad a=1,2 \,. \label{chargeLPFC}
\ee
Note here we need to identify the asymmetric factor in the complex SYK with the effective chemical potential $\bar\mu$ used in the WCFT through $\bar\mu \beta = - 2\pi \calE$.

Moreover, with such an identification the transformation law in (\ref{2pf1}) for the WCFT correlator under $ x\to  f(x)$, $y\to y+\Lambda(x)$ 
can be written as
\begin{align}
G_W(\tau_1,\tau_2;y_1,y_2)\rightarrow& ~~G'_W(\tau_1,\tau_2;y_1,y_2) \nn\\
&= \left( f'(\tau_1)\right)^{\Delta_1} \left( f'(\tau_2)\right)^{\Delta_2}
G_W(f(\tau_1),f(\tau_2);y_1+\Lambda(\tau_1),y_2+\Lambda(\tau_2)) \nn \\
&= \left( f'(\tau_1)\right)^{\Delta_1} \left( f'(\tau_2)\right)^{\Delta_2}
e^{i  (\Lambda(\tau_1)-\Lambda(\tau_2))}  G_W(f(\tau_1),f(\tau_2);y_1,y_2) \,,
\end{align}
and reduces to the transformation law in (\ref{GSfuncC}) for the complex SYK correlator.
Defining further the correlator of two fermionic operators with scaling dimension $(1-\Delta)$ and same charge, we can identify
\be \Sigma_W(\tau_1,\tau_2;y_1,y_2)\leftrightarrow e^{i(y_1-y_2)} \Sigma_c(\tau_1,\tau_2),\label{Smatch}\ee
and verify that not only the saddle point solutions are matched but also the transformation law from WCFT side matches the (\ref{GSfuncCS}) for the complex SYK.

To recapitulate, the first option for connecting  WCFT and the complex SYK  is to identify the Euclidean time $\tau$ in the complex SYK as the Virasoro direction in WCFT, and the phase as the $\UU(1)$ direction as shown in  (\ref{GwGcmatch}).
Also the tilting angle in the WCFT side is proportional to the asymmetric factor in the complex SYK by $\bar\mu\beta = -2\pi \mathcal{E}$.
 With the mapping  (\ref{GwGcmatch}), we relate the complex SYK Green's function to a WCFT correlation function for an operator with conformal weight $\Delta_W=\Delta$ and same $\UU(1)$ charge. At zero temperature with $\bar\mu\beta \equiv -2\pi \mathcal{E}$ fixed, one can also relate the zero temperature correlator (\ref{Gzero}) of the complex SYK model with that of WCFT in (\ref{2pfZero}).
 
 \end{defn}

\begin{defn}[Option II.]
Apart from the matching condition in (\ref{GwGcmatch}), one can also consider an alternative matching.  The Virasoro direction $x$ in WCFT can be viewed as a null direction such that, $x=y+\tau$ and here $\tau$ is once again the Euclidean time in the complex SYK model. By taking $y=0$, one can reproduce the results in the complex SYK.  In particular, the correlation functions are related by
\bea
G_{W}(\tau_1,\tau_2;0,0)=G_{c}(\tau_1,\tau_2) 
\,.
\label{GwGcmatch1}
\eea
One can check that two point function in WCFT (\ref{2pffiniteE}) once again reduces to the one in (\ref{Gfinite}) for the complex SYK model. Moreover, the transformation rules (\ref{2pf1}) under $ x\to  f(x)$,  $y\to y+\Lambda(x)$ 
are given  by
\begin{align}
G_W(\tau_1,\tau_2;0,0)\rightarrow& ~~G'_W(\tau_1,\tau_2;0,0) \nn\\
&= \left( f'(\tau_1)\right)^{\Delta_1} \left( f'(\tau_2)\right)^{\Delta_2}
G_W(f(\tau_1),f(\tau_2);\Lambda(\tau_1),\Lambda(\tau_2)) \nn \\
&= \left( f'(\tau_1)\right)^{\Delta_1} \left( f'(\tau_2)\right)^{\Delta_2}
e^{i(\Lambda(\tau_1)-\Lambda(\tau_2))}  G_W(f(\tau_1),f(\tau_2);0,0) \,,
\end{align}
which is identical to (\ref{GSfuncC}) using the identification \eqref{GwGcmatch1}. Physically, the second choice is to view the complex SYK model as living on the boundary of the 2d spacetime where the WCFT is defined. This is analogous to  the discussion of ZZ-boundary conditions \cite{Zamolodchikov:2001ah} of Liouville theory. The two different choices will not affect the discussions in this note, which is mainly kinematics  based on symmetries. In the field theory side, differences are expected to show up if we want to further match perturbations away from the IR fixed points, as the perturbation parameters will originate from different sources. In the holographic dual, option I corresponds to a usual Kaluza-Klein reduction along a  null $U(1)$  direction, as was discussed in \cite{Castro:2014ima, Cvetic:2016eiv}. Then the two dimensional theory will miss the massive modes from Kaluza-Klein tower. As will be seen in an upcoming paper \cite{Chaturvedi:2018x}, option II corresponds to, schematically,  a dimensional reduction along the $\phi\equiv {y\over\lambda }$ direction, where $\lambda$ is a parameter controlling how far away from the horizon and extremality. 
The two dimensional theories will therefore contain different subsets of the three dimensional phase space.  
We hope to report these further understandings in the future.

\end{defn}

\end{defn}

\subsection{Effective action and symmetry generators}
\label{section: effective action and symmetry generators}

\begin{defn}[The effective action in the complex SYK.]
Here we discuss the low temperature effective action and thermodynamics for the complex SYK model. The UV term $\partial_\tau$ in the complex SYK  explicitly breaks the time reparametrization symmetry $\tau\to f(\tau)$ and time dependent phase shift symmetry $\Lambda(\tau)$ of the action in Eq.(\ref{actioncsyk3}). In the low temperature regime, the effect of the explicit symmetry breaking can be summarized by an effective action that is only invariant under $\SL(2,\RR) \times \UU(1)$.  An effective action for infinitesimal diffeomorphisms was  proposed in \cite{Davison:2016ngz}, which can be naturally promoted to the  following form for finite diffeomorphisms:
\begin{align}
\frac{S_{\eff}}{N}=\frac{K}{2}\int^{\beta}_0 d\tau\left(\Lambda'(\tau)-i \bar{\mu} f'(\tau)\right)^2-\frac{\gamma}{4\pi^2}\int^{\beta}_0 d\tau\left\{\tan\left(\frac{\pi f(\tau)}{\beta}\right),\tau\right\}\,,\label{csykSeff}
\end{align} where $K$ is the compressibility and $\gamma$ is the specific heat.
This action for the finite diffeomorphisms is consistent with the one proposed in \cite{Davison:2016ngz} upto a constant (independent of $f(\tau)$ and $\Lambda(\tau)$) and total derivative terms.
 This effective action is natural from the symmetry breaking pattern that will be discussed later. The same form is also noted in
\cite{Gaikwad:2018dfc} using a holographic approach.
\end{defn}

\begin{defn}[Quantization conditions v.s. spatial circles.]
In general, we can define a Hilbert space on any ``spatial circle'' specified by $(\tau,y)\sim (\tau+2\pi a i, y+2\pi \baa)$, which we will denote as $(a i,\baa)$ hereafter.
The above periodicity can be realized by a conformal map
\begin{align}
z = e^{\frac{\tau}{a}}, \quad \bar{z} = y + i \frac{\bar{a}}{a} \tau \,.
\end{align}
 The Noether charges \footnote{Note that there is an apparent sign difference between our charges and those in \cite{Detournay:2012pc} and \cite{Castro:2015csg}. We differ from \cite{Detournay:2012pc} by an overall minus sign of the Kac-Moody generators. On the cylinder, we have $l_0=i\p_\phi,\,p_0=-i\p_{\bar z}$, while  \cite{Detournay:2012pc} has an opposite relative sign, see the below eq. (10) of \cite{Detournay:2012pc}. In \cite{Castro:2015csg}, the cylinder is defined by $z=e^{-\phi}$. In any case, it is easy to check that the charge definition is compatible with the generators \eqref{wcftgen} and the transformation rules \eqref{eqn: finiteTrans}. } can be written as
 \begin{align}
 L_n^{(ai,\baa)}= -\frac{i}{2\pi}  \int_0^{2\pi ai} d\tau\, e^{\frac{n\tau}{a} } \,T(\tau),\quad P_n^{(ai,\baa)}= \frac{1}{2\pi} \int_0^{2\pi ai} d\tau\, e^{\frac{n\tau}{a}} \,P(\tau),\label{lpaab}
\end{align}
where  the superscript  $(ai,\bar a)$ denotes the spatial circle and hence the quantization condition. Following \cite{Castro:2015csg}, we denote the spatial circle with  $(i,0)$ as the {\it canonical circle}.\footnote{Similar to the discussion in the previous section, here again we have two options, identifying $y \leftrightarrow\tau$, or as $y \leftrightarrow y+t\to y-i\tau$ with $y\to0$ in the final result. In the main text, we keep the first choice. For the second choice, discussion about charges and modular properties are similar to that in \cite{Castro:2015uaa}, where the canonical spatial circle is $(1,-1)$ as pointed in section-3 of \cite{Castro:2015uaa}.} One can check that charges on the {\it canonical circle} satisfy the same Virasoro-Kac-Moody algebra defined in (\ref{wcftVU1}). Note that under the transformation, $\tau'={\tau\over a}$ and  $\bar{z}' =y+i {\baa\over a} \tau$  one can rewrite  charges on the {\it canonical circle} in terms of the charges on arbitrary circle as
\be
L_n^{cl}\equiv L_n^{(i,0)}=a L_n^{(ai,\baa)}-\baa P_n^{(ai,\baa)}+{k\over 4}\baa^2 \delta_{n,0},\quad P_n^{cl}\equiv P_n^{(i,0)}= P_n^{(ai,\baa)}-{k\over 2}\bar a \delta_{n,0},\label{lpaab1}
\ee
where the superscript $cl$ denotes the charges on the cylinder with spatial circle parameterized by the {\it canonical circle}.
We use the notation $(ai,\bar{a})$ to emphasize that the spatial circle we choose is along the real time direction, i.e. $a \in \RR$. In contrast, the periodicity for imaginary time $\tau \rightarrow \tau + \beta$ will be referred as ``thermal circle'' following the terminology of \cite{Castro:2015csg}.
\end{defn}

\begin{defn}[Virasoro-Kac-Moody generators.]
We consider a tilted cylinder geometry generated by the following transformation
\begin{align}
~& z= \exp \left(\frac{2\pi i}{\beta} f(\tau)\right),   \quad \bar z=y
+
(\Lambda(\tau)-i\bar\mu f(\tau) ),\label{wcfttrans2}
\end{align}
where $f(\tau)$ and $\Lambda(\tau)$ are two arbitrary functions defined on the thermal circle satisfying the boundary conditions
\begin{align}
 f(\beta)=f(0)+\beta, \quad  \Lambda(\beta)=\Lambda(0),
 \label{bcfL}
\end{align}
i.e. we only consider the winding number $1$ reparametrizations of time and winding number $0$ phase fluctuations.
 In the new coordinates $(\tau,\,y)$, we consider states quantized with via the boundary conditions in (\ref{bcfL}).
 Using the terminology introduced in our previous subsection, we are effectively treating
 the apparent ``thermal circle''  $(\tau,y)\sim(\tau+ \beta,\,y {+} i \beta \bar{\mu})$ as a ``spatial circle'', in the  sense that the quantization condition is defined on that circle, and charges will be written as an integral along the circle. It might seem confusing here why we change a ``thermal circle '' to a ``spatial circle''. As we will discuss later, swapping the circles is the essential step of the S-transformation of the torus partition function of WCFT.

 By employing the transformation rules in \eqref{eqn: finiteTrans}, the stress energy tensor and $\operatorname{U}(1)$ Kac-Moody current on the tilted thermal cylinder are given as
\begin{align}
P(\tau) &
=
{-}
\frac{k}{2} \left(\Lambda'(\tau)-  i \bar{\mu}  f'(\tau) \right) \,,
\label{Pp3} \\
T(\tau)
&= \frac{c}{12} \left\{ \tan \frac{\pi}{\beta} f(\tau),\tau \right\} - \frac{k}{4} \left( \Lambda'(\tau)- i \bar{\mu} f'(\tau) \right)^2
\,,
 \label{Tp3}
\end{align}
where we have assumed that the plane vacuum has zero charges.
From the definition of the charges in (\ref{lpaab}), using the above expressions for the currents $P(\tau)$ and $T(\tau)$ one can obtain the Virasoro-Kac-Moody charges on the quantization along the circle with periodicities $(a'i,\bar a')=(\beta/2\pi,i \beta \bar{\mu}/2\pi)$ as 
\begin{align}
P_0^{(a'i,\bar a')}&=i{k\beta\bar{\mu}\over4\pi},\label{thermalP0}\\
L_0^{(a'i,\bar a')}&=-\frac{i c}{24\pi}  \int_0^{\beta} d\tau\, \left(  \left\{ \tan \frac{\pi}{\beta} f(\tau),\tau \right\} - \frac{3k}{c} \left( \Lambda'(\tau)- i \bar{\mu} f'(\tau) \right)^2  \right)
\,.
\label{thermalL0}
\end{align}
Moreover, using (\ref{lpaab1}) we obtain the zero modes of the charges on the cylinder with {\it canonical circle} as
\begin{align}
P^{cl}_0 &=0 \,,
\label{Pp4}\\
L^{cl}_0 &=- \frac{c\beta}{48\pi^2} \int_0^{\beta} d\tau \left(  \left\{ \tan \frac{\pi}{\beta} f(\tau),\tau \right\} - \frac{3k}{c} \left( \Lambda'(\tau)- i \bar{\mu} f'(\tau) \right)^2  \right)+{k\beta^2\bar\mu^2\over 16\pi^2}.\label{Tp4}
\end{align}
Note that the global properties of the thermal cylinder do not depend on the functions $f(\tau)$ and $\Lambda(\tau)$ as long as they satisfy the boundary conditions \eqref{bcfL}. The expectation value of the Kac-Moody zero modes  $P_0^{cl}$ does not depend on the choice of the arbitrary functions $f(\tau),\,\Lambda(\tau), $ while the higher Kac-Moody modes $P_n^{cl}$ do.
Generically, $L_0^{cl}$ depends on $f(\tau)$ and $\Lambda(\tau)$, except for the zero modes, namely $\SL(2,\mathbb{R})\times \UU(1)$ transformation \eqref{sl2tr} \eqref{u1tr}.
\end{defn}

\begin{defn}[The matching.]

Interestingly, it can be observed that with the identification
\be
{c\over k}={3\gamma\over 2\pi^2 K},\label{ckratio}
\ee
the effective action (\ref{csykSeff}) for the complex SYK model in the IR limit  can be written in terms of the charges of the warped CFT as\footnote{the analogous comparison between the Virasoro charge of a large $c$ 2d CFT and the effective action in the Majorana SYK was observed in \cite{Mertens:2017mtv}.
}
\begin{align}
S_{\eff}&= - \frac{N\gamma}{4\pi^2} \int_0^{\beta} d\tau \left(  \left\{ \tan \frac{\pi}{\beta} f(\tau),\tau \right\} - \frac{3k}{c} \left( \Lambda'(\tau)- i \bar{\mu} f'(\tau) \right)^2  \right)\nn\\
&=-i\frac{6 N \gamma }{\pi c} L_0^{(a'i,\bar a')}
=\frac{12 N \gamma }{\beta c} \left(L_0^{cl}-\frac{k\beta^2\bar\mu^2}{16\pi^2}\right),
\label{Seffmatch}
\end{align}
where $L^{cl}_0$ is the zero mode of the Virasoro generators on the cylinder with ``spatial circle'' parameterized by the {\it canonical circle}. 
\end{defn}

\subsection{Partition function and modular covariance}

In this subsection, we discuss partition function and
 the modular properties of the WCFT. The modular covariance of the WCFT naturally explains the apparent puzzling quantization on thermal circle in last subsection. Furthermore, by evaluating the partition function in the WCFT we can match both the classical piece of the complex SYK free energy and its quantum fluctuations, i.e. logarithmic corrections.

\begin{defn}[Partition function and thermodynamics for the complex SYK model.]
The partition function for the effective action of SYK can be written as a path integral over the reparamterizations and phase fluctuations:
\begin{align}
Z = \int \frac{\calD f \calD \Lambda }{\SL(2,\RR)\times \UU(1)} \exp \left[ N \int_0^{\beta} d\tau \frac{\gamma}{4\pi^2} \left\lbrace \tan  \frac{\pi}{\beta}f(\tau)  ,\tau \right\rbrace - \frac{K}{2} \left( \Lambda'(\tau) - i\bar{\mu} f'(\tau)  \right)^2   \right] \,.
\label{eqn: partition function SYK}
\end{align}
Considering the leading saddle point  $\Lambda(\tau)=\const$ and $f(\tau)=\tau$, the large $N$ grand potential from the effective action (\ref{csykSeff}) is given by
\begin{align}
\Omega_{\eff}:=
\Omega- E_0 +  N \frac{ s_0}{\beta}
=- \frac{N}{2}  \left(  K \bar\mu^2+\frac{\gamma}{\beta^2} \right) + \ldots ,  \label{csykSeff1}
\end{align}
where $E_0$ is the ground state energy and $Ns_0$ is the zero temperature entropy\cite{Sachdev:2015efa}. Neither of these terms is described by the effective action (\ref{csykSeff}) and we will denote the left hand by $\Omega_{\eff}$ for convenience. 
From the classical piece of the grand potential defined by $\Omega= E- \frac{1}{\beta} S - \mu Q$,
we can extract the thermodynamic quantities:
\begin{align}
{S}= {\beta^2} {\delta \Omega_{\eff} \over \delta \beta}=N{\gamma\over\beta},\quad
Q=
-{\delta \Omega_{\eff}\over \delta \bar\mu}=NK\bar\mu,\nn\\
{E}= \Omega_{\eff} + \frac{S}{\beta} + \bar{\mu} Q=\frac{N}{2} \left(K \bar\mu^2 + \frac{\gamma }{\beta ^2} \right), \label{csykcharges}
\end{align}
where $S$ is the entropy (excluding the zero temperature entropy), $E$ is the energy above ground state energy and $Q$ is the total charge. Using the above relations one can rewrite the formula for entropy
\begin{align}
S= 2\pi\sqrt{{N\gamma\over2\pi^2} \left(E- {Q^2\over 2NK}\right)}  \,,
\label{csyks}
\end{align}
in the microcanonical ensemble.  Here we emphasize that the entropy $S$, charge $Q$ and energy $E$ are  obtained  from the effective grand potential $\Omega_{\eff}$ which excludes the zero temperature entropy and ground state energy.

Beyond the saddle point, there is a logarithmic correction term\cite{Davison:2016ngz}:
\begin{align}
\Delta \Omega_{\eff}  = \frac{2}{\beta} \log \beta J \,.
\end{align}
Comparing to the correction for the Majorana SYK, which is $\frac{3}{2\beta}\log \beta J $ and 1-loop exact\cite{Stanford:2017thb}, the additional correction here is $\frac{1}{2\beta} \log \beta J$. It comes from the $\UU(1)$ fluctuation
\begin{align}
\frac{S^{\UU(1)}_{\eff}}{N}=\frac{K}{2}\int^{\beta}_0 d\tau\left(\Lambda'(\tau)-i \bar{\mu} f'(\tau)\right)^2,
\end{align}
excluding the zero mode $\Lambda \rightarrow \Lambda +\const$.
For $\bar{\mu}=0$, the 1-loop correction is obviously exact. Moreover, for non-zero $\bar{\mu}$ we can redefine the field $\tilde{\Lambda}(\tau) := \Lambda(\tau) - i\bar{\mu} f(\tau)$ to absorb the mixing between the reparametrization and phase fluctuation. The redefinition does not change the integral measure in the path integral and only affects the boundary condition $\tilde{\Lambda}(\beta)=\tilde{\Lambda}(0)- i \bar{\mu} \beta $. Therefore the 1-loop calculation for the $\UU(1)$ fluctuation is also exact.
\end{defn}

\begin{defn}[Modular covariance in WCFT and thermodynamics.]
In \cite{Detournay:2012pc, Castro:2015csg}, a Cardy like formula for the asymptotic density of states of a WCFT was derived using the Virasoro-Kac-Moody symmetries.
This involves finding a transformation of the form (\ref{wcftTrans}) which exchanges a thermal cycle with a spatial cycle and plays the role of a modular transformation. In Euclidean coordinates \footnote{Note that comparing to \cite{Detournay:2012pc, Castro:2015csg}, we are rederiving the formula in Euclidean space, where the Virasoro direction is viewed as the time direction. More explicitly, the charges are defined as in (\ref{lpaab}).}, considering the WCFT on a torus with
\begin{align}
\text{thermal circle} \quad  (\tau,y) &\sim (\tau+2\pi b, y+2\pi \bar b),\label{Tcircle} \\
\text{spatial circle} \quad (\tau,y)&\sim (\tau+2\pi a i ,y+2\pi \bar{a}),\label{Scircle}
\end{align}
the partition function for the WCFT can be written as
\begin{align}
Z_{(ai|\bar{a})}(b|\bar{b})&=\Tr_{(ai| \bar{a})} e^{-2\pi b L_0+2\pi i\bar{b} P_0}\label{first}\\
&=e^{i\pi k \bar{a} \left(\bar{b}+{ib\bar{a}\over2a}\right)}Z_{(i|0)}\left(\left.{b\over a}\right\vert \bar{b}+i{\bar a\over a}b\right)\label{second}\\
&=e^{i\pi k \bar{a} \left(\bar{b}+i{b\bar{a}\over2a}\right)}Z_{\left(\left.{b\over  a}\right\vert \bar{b}+i{\bar a\over a}b\right)}(-i|0)  \label{third}\\
&= \Tr_{(i|0)}\exp\left(-{\pi k a\bar{b}^2\over 2 b}
+
2\pi i \left(i{a \bar{b}\over b}-\bar{a}\right)P^{cl}_0-{2\pi a \over b}L^{cl}_0 \right) , \label{fourth}
\end{align}
where the trace   
is taken over the Hilbert space with quantization condition specified by the subscript.
 The first line (\ref{first}) is the definition of the partition function for a theory on spatial circle $(ai,\bar{a})$ and evolved along the thermal circle parameterized by $b,\bar b$. In \eqref{second}, the warped conformal transformation $\tau \to {\tau\over a}$ and  $y\to y+i{\bar a\over a}\tau$, is employed to rewrite the partition function on the canonical spatial circle $(\tau,y)\sim(\tau+2\pi i,y)$. While in the third line \eqref{third} the canonical spatial circle is swapped with the thermal circle, an operation which is valid in any quantum field theory. In the fourth line \eqref{fourth} the warped conformal transformation $\tau\to  i{a\over b}\tau$ and $y\to y+i(i{a\bar b\over b}-{\bar a}) \tau$, brings the theory back to the canonical circle. 

Finally, we take the vacuum character:
\begin{align}
\chi_{\vac} = \Tr^{\vac}_{(i|0)}\exp\left(-{\pi k a\bar{b}^2\over 2 b}
+
2\pi i \left(i{a \bar{b}\over b}-\bar{a}\right)P^{cl}_0-{2\pi a \over b}L^{cl}_0 \right) \,,
\label{wcftZ}
\end{align}
the trace over all states will become the trace over the vacuum module which we denote as ``$\Tr^{\vac}$''.  The identity module of the WCFT \eqref{wcftZ} can be evaluated as
\begin{align}
\chi_{\vac} &= \exp\left(-{\pi k a\bar{b}^2\over 2 b} +2\pi i \left(i{a \bar{b}\over b}-\bar{a}\right)P^{\vac}_0-{2\pi a \over b}L^{\vac}_0 \right) \chi^{des}_{\vac}
\nonumber\\
&= \exp\left(\underbrace{{\pi a c\over12 b}-{\pi k a\bar{b}^2\over 2 b} + 2\pi i \left(i{a \bar{b}\over b}-\bar{a}\right)P_0^{\vac}-{2\pi a(P^{\vac}_0)^2\over b k}}_{=:-I_c\quad \text{classical contribution}} \right) \chi^{des}_{\vac},
\label{app}
\end{align}
where in the above expressions, 
$L_0^{\vac}$ and $P_0^{\vac}$ are the expectation values of the charges defined on the canonical circle. In the second line of the above expressions we have used the condition for the vacuum, $L_0^{\vac}=-{c\over 24}+{(P_0^{\vac})^2\over k}$ \cite{Detournay:2012pc}.  We have also split the partition function into a classical contribution and a correction from descendants denoted by $\chi_{\vac}^{des}$.

Let us first consider the classical piece. With $b$ and $\bar b$ as the thermodynamic periods, one can obtain a Cardy like formula for the thermal entropy of the WCFT:
\begin{align}
S=-(1-b\p_b-\bar b\p_{\bar b})\,I_{c}
={\pi a c\over6 b}-{4\pi a (P^{\vac}_0)^2 \over b k}+2\pi i \left(i{a \bar{b}\over b}-\bar{a}\right)P_0^{\vac}.
\end{align}
We also have
\bea
\left\langle P_0\right\rangle=-{1\over 2\pi i}\p_{\bar b}\,I_{c}=i{ a\over b} \left(P_0^{\vac}+{k\bar b\over2 }\right),
\quad
\left\langle L_0\right\rangle={1\over 2\pi} \p_b \,I_{c}
={a\over b^2} {c\over 24}+{\left\langle P_0\right\rangle^2\over a k}. \label{l0vev}
\eea
Following \cite{Detournay:2012pc}, we express the entropy as $S= S_L+S_R$ \footnote{
 According to the original argument in \cite{Detournay:2012pc}, the entropy formula applies to WCFTs with a discrete spectrum with $L_0$ bounded from below, and at the large temperature or equivalently, in the ${b\over a}\to  0$ region.
We expect it also to be true if the spectrum of the WCFT is sparse and the central charge is large, along the lines of argument of \cite{Hartman:2014oaa}. }
\begin{align}
S_L=-2\pi i{\bar a} P_0^{\vac}+ \frac{4\pi i \left\langle P_0\right\rangle P_0^{\vac}}{k},\quad
S_R=2\pi\sqrt{\frac{c}{6}\left(a \left\langle L_0\right\rangle-\frac{\left\langle P_0\right\rangle^2}{k}\right)}.\label{n18}
\end{align}
For WCFT with a neutral vacuum, $S_L=0$, and the entropy only comes from $S_R$.  $S_L$ contains an explicit $i$, but it becomes real when the vacuum charge is purely imaginary. As was shown in explicit examples in \cite{Detournay:2012pc}, and from the modular bootstrap in \cite{Apolo:2018eky},  holographic WCFT with a negative level does feature a pure imaginary $P_{\vac}$.

The $\chi^{des}_{\vac}$ term denotes the contribution from the Virasoro-Kac-Moody descendants in the vacuum character, which can be calculated by a method discussed in \cite{Compere:2013bya}. More explicitly, we construct all the states that are generated by acting $L_{-n}, P_{-m}$ with positive integers $n$ and $m$ on the vacuum state. Notice that the vacuum is invariant under $\operatorname{SL}(2,\mathbb{R}) \times \operatorname{U}(1)$ and generic descendant states can be written as linear combinations of the generic state
\begin{align}
\prod_{n \geq 2,\, m\geq 1} L_{-n}^{N_n}P_{-m}^{N_m}|0\rangle\,,
\end{align}
where $\{ N_n \}$ and $\{ N_m \} $ are integer sequences.
Note that we start from $n=2$ for Virasoro generators, since $L_{-1}$ annihilates the vacuum. Thus the contribution from the descendants \footnote{Here we will only consider WCFT theories with $c\ge$2, so that there are no  additional null states in the vacuum module.}  is given by
\begin{align}
\chi^{des}_{\vac}=
\frac{
\Tr \left(q^{L_0} \bar q^{P_0}\right)}
{q^{L^{\vac}_0}  \bar q^{P^{\vac}_0}}
= \prod_{n\geq 2,\, m\geq 1} \sum_{N_n,N_m\geq 0} \, q^{nN_n+mN_m}
=\frac{q^{1\over 12}\,(1-q)}{\eta(\tau)^2} \,,\label{wcftchar}
\end{align}
where  $q=e^{2\pi i \tau}$ and $\bar q=e^{2\pi i \bar \tau}$ and with abuse of notation we have identified  $\tau= i \frac{a}{b}$ and $\bar \tau = \left(i{a \bar{b}\over b}-\bar a\right)$ to be the modular parameters defining the torus for the WCFT. 
$\eta(\tau)$  is the Dedekind eta function. In the $q\rightarrow 1$ limit, we can use the modular properties of the eta function to find
\begin{align}
\eta(\tau)^2 = \frac{1}{-i\tau} \eta \left(-\frac{1}{\tau}\right)^2  \approx \frac{1}{-i\tau}e^{-{\pi i\over 6\tau}}=
\frac{b}{a} e^{- \frac{\pi b}{6a}} \quad \Rightarrow \quad \chi^{des}_{\vac}\approx   2\pi \left({a\over b}\right)^{2}e^{\frac{\pi b}{6a}} \,,
\label{modeta}
\end{align}
where we have used the series expansion of the Euler function $\phi(q) =\sum_{n\in \ZZ} (-1)^n q^{(3n^2-n)/2}$ for the second step, which is related to the Dedekind's eta function by $\eta(q)=q^{1\over24}\phi(q)$.
The ${b\over a}$ term can be absorbed into the ground state energy.

\begin{comment}
Therefore, we can can split the effective action into a classical piece and a quantum fluctuation piece, i.e. the contribution from the character:
\begin{align}
-\log Z \approx \underbrace{-2\pi i \left(i{a \bar{b}\over b}-\bar{a}\right)P_0^{\vac}+{2\pi a(P^{\vac}_0)^2\over b k} -{\pi a c\over12 b}+{\pi k a\bar{b}^2\over 2 b} 
}_{=:I_c\quad \text{classical contribution}} \nn \\
\underbrace{-\log\left[2\pi \left({a\over b}\right)^2\right] 
\textcolor{red}{- \frac{\pi b}{6a}}
 }_{\text{fluctuations/descendants}}
\label{eqn: WCFT free energy}
\end{align}
\YG{\bf [I added last factor back: "shift of ground state energy", shall we comment on that?]}{}
\end{comment}

\end{defn}

\begin{defn}[The matching.]
When the vacuum dominates, the WCFT partition on a torus is given by (\ref{wcftZ}) where the trace is over the identity module.
Then one immediately recognizes the similarity between the WCFT partition function and the complex SYK partition function \eqref{eqn: partition function SYK}.
The expression (\ref{wcftZ}) for the WCFT partition function is general, while applying it to the connection to the complex SYK model, we need to fix some parameters.
In particular, we are interested in the complex SYK model on a thermal circle tilted by a chemical potential:
\begin{align} 
b= \frac{\beta}{2\pi} , \quad \bar{b} = \frac{i \bar{\mu} \beta }{2 \pi} \, .
\label{thermalchoice}
\end{align}
We also introduced an extra spatial circle parameterized by $(ai,\bar a)$ where the Hilbert space is defined on.
If we identify 
\be 
{c}={3N\gamma \over \pi^2 a},\quad {c\over k}={3\gamma\over 2\pi^2 K},\quad P_0^{cl}=P_0^{\vac}=0 \,,
\label{paraanstz}
\ee
the 
vacuum character \eqref{wcftZ} matches the complex SYK partition function (\ref{eqn: partition function SYK}). 
Alternatively, one can follow the argument of \cite{Mertens:2017mtv} to relate \eqref{wcftZ} with \eqref{eqn: partition function SYK}. The path integral in \eqref{eqn: partition function SYK}  can be understood as the coadjoint orbit of ${\operatorname{Diff}(S^1)\times \UU(1)_{\text{local}} \over \SL(2,\mathbb{R})\times \UU(1)}$. In the semiclassical limit $c\rightarrow \infty$ and $q\rightarrow 1$, the path integral is expected to become the trace over  the vacuum module \eqref{wcftZ}. 

Now let us use the modular transformation outlined in \eqref{first} to \eqref{wcftZ} to reinterpret the 
matching of the complex SYK partition function and the WCFT charge defined on the thermal cycle in section~\ref{section: effective action and symmetry generators}.
We can start with an arbitrary torus, identify a spatial circle and evolve along the thermal circle
$b= \frac{\beta}{2\pi}$, $\bar{b} =  \frac{i \bar{\mu} \beta }{2 \pi}$.
 But a rescaling will bring us to a canonical circle as shown in \eqref{second}. 
From \eqref{second} to \eqref{third}, we swap the spatial circle and the thermal circle. More explicitly,
in \eqref{third}, states are quantized with periodicity specified by the previous thermal circle, and evolved along a circle with $(-i,0)$.
Swapping the circles explains why we wrote down the charges \eqref{thermalP0} and \eqref{thermalL0} on the thermal circle in section~\ref{section: effective action and symmetry generators}.
 In this note, we keep $a$ and $\bar a$ as free parameters which is sufficient for our purposes.

One more comment is about the vacuum charge $P_0^{\vac}$ which we set to $0$ in this note, whereas a more general WCFT may contain a non-zero vacuum charge. This charge will contribute to the entropy at the zero temperature limit. On the complex SYK side, there is also a non-vanishing zero temperature entropy. It is possible that there are further connections which we would like to explore in the future.

With the parameters \eqref{thermalchoice} and \eqref{paraanstz} fixed, one can also compare the thermodynamical quantities and the 1-loop corrections. 
We first match the thermodynamical quantities and in particular the formula for entropy\footnote{The $-$ sign in the charge formula \eqref{match charge} is a consequence of the convention: we define a positive phase shift for the first operator appearing in the correlation function, which is an annihilation operator in the definition of the charge $Q$.}
\begin{align}
&\langle P_0 \rangle|_ {P_0^{\vac}=0} = i \frac{ak \bar{b}}{2b} = {-}  NK\bar{\mu} 
= {-}   Q   \,, \label{match charge}
 \\
&\langle L_0\rangle|_ {P_0^{\vac}=0} = \frac{ac}{24b^2} + \frac{\langle P_0 \rangle^2}{ak} =  \frac{N}{2} \left( K \bar{\mu}^2 + \frac{\gamma}{\beta^2} \right)
=  E  \,,
\\
&S_R = 2\pi \sqrt{  \frac{c}{6} \left( a\langle L_0 \rangle - \frac{\langle P_0 \rangle^2}{k}  \right) } = 2\pi \sqrt{\frac{N\gamma}{2\pi^2} \left( E- \frac{Q^2}{2NK} \right) } = S \,.
\end{align}
Note that all the quantities on the left hand side are calculated from the effective action and hence have already subtracted the zero-temperature part.  Furthermore, the contribution to the effective action from WCFT descendants
as shown in \eqref{modeta}
 using the complex SYK parameters is given by
%\wei{\begin{align}
%-\log \left[ 2\pi \left( \frac{a}{b} \right)^2 \right]= 2 \log \frac{\beta}{a} \textcolor{red}{- \frac{\pi b}{6a}}= 2\log \beta \textcolor{red}{ - \frac{\beta}{12 a}}+ \text{($\beta$ independent terms)}.
%\end{align}}{If we just write the log term, we don't need to mention the $b/a$ term}
\begin{align}
-\log \left[ 2\pi \left( \frac{a}{b} \right)^2 \right]= 2 \log \frac{\beta}{a} = 2\log \beta + \text{($\beta$ independent terms)}.
\end{align}
The $2\log \beta$ term agrees with the 1-loop correction in the complex SYK.%  \YG{$-\frac{1}{12a}$ is the shift of the ground state energy.}{Any further explanation?}

\end{defn}

\begin{defn}[A side comment from holography.]
Holographically, WCFT has been proposed to describe quantum gravity at the near horizon throat of extremal black holes. In particular,  the phase space of Einstein gravity on asymptotically AdS$_3$ with the CSS boundary conditions \cite{Compere:2013bya} is organized by the warped conformal symmetry. 
It has been shown \cite{Detournay:2012pc,Compere:2013bya,Song:2016gtd} that the entropy of a holographic WCFT \eqref{n18} reproduces the entropy of warped AdS$_3$ black holes as well BTZ black holes with the CSS boundary conditions. 
Furthermore, the descendant contribution to the vacuum character \eqref{wcftchar} also agrees with the 1-loop determinant on BTZ black holes in the bulk, see eq. (3.31) of \cite{Castro:2017mfj}.  These results indicate that the vacuum character \eqref{wcftZ} is indeed the relevant part for black hole physics. 
 \end{defn}

\section*{Acknowledgements}
We thank Dionysios Anninos, 
Luis Apolo, 
Alex Belin, Alejandra Castro, 
Mirjam Cvetic,  
Monica Guica, 
Finn Larsen, 
Juan Maldacena,
Ioannis Papadimitriou, 
Xiao-Liang Qi,
Subir Sachdev,
Steve Shenker, 
Andrew Strominger, 
Herman Verlinde, 
Jianfei Xu, 
Hong Yao, 
Hui Zhai for helpful discussions.  The work of PC, WS, and BY was supported by the National Thousand-Young-Talents Program of China, and NFSC Grant No. 11735001. YG is supported by the Gordon and Betty Moore Foundation EPiQS Initiative through Grant (GBMF-4306). All the authors thank the Tsinghua Sanya International Mathematics Forum
for hospitality during the workshop and research-in-team program on ``Black holes, Quantum
Chaos, and Solvable Quantum Systems''.  WS thanks the Solvay workshop on ``Holography'' on which she presented the work and received helpful feedback.

\bibliographystyle{JHEP}

\bibliography{ref.bib}

\end{document}